\documentclass[twoside,12pt]{article}
\usepackage{epsfig}
\def\Journal#1#2#3#4{{#1} {#2} (#4) #3 }

\def\NPA{{\em Nucl. Phys.} A}

\def\PLB{{\em Phys. Lett.} B}

\def\PRL{\em Phys. Rev. Lett.}

\def\PREP{\em Phys. Rep.}

\def\PRC{{\em Phys. Rev.} C}

\def\ZPA{{\em Z. Phys.} A}
\def\JPG{{\em J. Phys.} G}
\def\EPJA{{\em Eur. J. Phys.} A}

\def\ANN{\em Annu. Rev. Nucl. Part. Sci.}
\def\SCI{\em Science}

\begin{document}
\newcommand{\p}{\partial}
\newcommand{\ls}{\left(}
\newcommand{\rs}{\right)}
\newcommand{\beq}{\begin{equation}}
\newcommand{\eeq}{\end{equation}}
\newcommand{\beqa}{\begin{eqnarray}}
\newcommand{\eeqa}{\end{eqnarray}}
\newcommand{\bdm}{\begin{displaymath}}
\newcommand{\edm}{\end{displaymath}}
\newcommand{\fps}{f_{\pi}^2 }
\newcommand{\mks}{m_{{\mathrm K}}^2 }
\newcommand{\ms}{m_{{\mathrm K}}^{*} }
\newcommand{\msq}{m_{{\mathrm K}}^{*2} }
\newcommand{\rhos}{\rho_{\mathrm s} }
\newcommand{\rhob}{\rho_{\mathrm B} }
\title{\vspace{1cm} Strangeness production in heavy ion reactions 
at intermediate energies}
\author{Christian Fuchs\\
Institut f\"ur Theoretische Physik\\ 
Universit\"at T\"ubingen, D-72076 T\"ubingen, Germany }
\date{}
\maketitle
\begin{abstract}
Kaon production, in particular $K^+$ production in heavy ion collisions at 
intermediate energies is discussed. Main emphasis is put on the 
question if subthreshold  $K^+$ production can serve as a
suitable tool to test the high density phase of such reactions and 
to deliver information on the high density behavior of the nuclear 
equation of state. It is shown that  
the  $K^+$  excitation function in heavy ($Au+Au$) over 
light ($C+C$) systems provides a robust observable which, 
by comparison to data, strongly favors a soft equation of state. 
A second question of interest is the existence of an in-medium kaon
potential as predicted by effective chiral Lagrangiens. Here it 
is argued that transport calculations support this scenario 
with, in the meantime, a significant level of consistency.
\end{abstract}

\section{Introduction}
The original motivation to study the kaon production 
in heavy ion reactions at intermediate energies, namely to extract 
information on the nuclear equation of state (EOS) at high densities is 
a matter of current debate. Already in the first theoretical 
investigations by transport models it was noticed that the $K^+$ 
yield reacts sensitive on the nuclear equation of state 
\cite{AiKo85,qmd93,li95b,baoli94}. The yields were 
found to be about a factor 2--3 larger when a 
soft EOS was applied compared 
to a hard EOS. At that time the available data \cite{kaos94} 
already favored a soft equation of state. However, calculations 
as well as the experimental data were still burdened with 
large uncertainties. 

In \cite{fuchs01} we studied the question if in the meantime 
decisive information on the nuclear EOS can be extracted from subthreshold 
kaon production in heavy ion collisions. There are several reasons 
why it appears worthwhile to do this: Firstly, there has been 
significant progress in the recent years towards a more precise 
determination of the elementary kaon production cross sections 
\cite{sibirtsev95,tsushima99}, based also on new data points from 
the COSY-11 for the reactions 
$pp \longrightarrow p \Lambda K^+ $ very close to threshold \cite{cosy11}. 
Secondly, the KaoS Collaboration has performed systematic 
measurements of the $ K^+$ production far below threshold in 
heavy ($Au+Au$) and light ($C+C$) systems \cite{sturm00}. Looking at the 
ratios built from heavy and light systems possible uncertainties which 
might still exist in the theoretical calculations should cancel out 
to a large extent which allows to draw reliable conclusions. 
Furthermore, far below threshold the kaon production is a highly 
collective process and a particular sensitivity to the compression of the 
participant matter is expected. 

\section{The Model}
The present investigations are based on the 
Quantum Molecular Dynamics (QMD) transport model \cite{ai91}. 
For the nuclear EOS we adopt soft and hard Skyrme forces 
corresponding to a compression modulus of 
K=200 MeV and 380 MeV, respectively, and with a momentum dependence 
adjusted to the empirical optical nucleon-nucleus potential 
\cite{ai91}. The saturation 
point of nuclear matter is thereby fixed at $E_B = -16$ MeV and 
$\rho_{\rm sat}= 0.17~{\rm fm}^{-3}$ \cite{ai91}. 
The calculations include $\Delta (1232)$ and $N^*(1440)$ resonances. 
The QMD approach with Skyrme interactions is well tested, 
contains a controlled momentum dependence and provides 
a reliable description of the reaction dynamics in the SIS 
energy range, expressed e.g. by collective nucleon flow observables as well 
as particle production. In contrast to AGS energies where 
the creation of resonance matter may lead to an effective 
softening of the EOS, baryonic resonances with masses 
above the $N^*(1440)$ can safely be neglected for the reaction 
dynamics at SIS energies.

We further consider the influence of an 
in-medium kaon potential based on 
effective chiral models \cite{kapla86,weise93,li95,brown962,lutz98}. 
The $K^+$ mean field consists of a repulsive vector part 
$V_\mu = 3/8 f^{*2}_{\pi} j_\mu$ and an attractive scalar part 
$\Sigma_S = m_{{\rm K}} -  m_{{\rm K}}^* = m_{{\rm K}} - 
\sqrt{ m_{{\rm K}}^2 - \Sigma_{\mathrm{KN}}/f_{\pi}^2\rho_S 
     + V_\mu V^\mu }$. Here $j_\mu$ is the baryon vector current 
and $\rho_S$ the scalar baryon density and  
$\Sigma_{\mathrm{KN}} = 450$ MeV. Following \cite{brown962} in 
the vector field the pion decay constant in the medium 
$f^{*2}_\pi = 0.6 f^{2}_\pi$ is used. However, 
the enhancement of the scalar part using $f^{*2}_\pi$ is compensated 
by higher order contributions in the chiral expansion \cite{brown962}, 
and therefore here the bare value is used, i.e. 
$\Sigma_{\mathrm{KN}}\rho_S /f^{2}_\pi$.   
Compared to other chiral approaches \cite{weise93,li95} the 
resulting kaon dispersion relation shows a relatively strong 
density dependence. The increase of the 
in-medium $K^+$ mass ${\tilde m}_{\mathrm K}$, 
Eq. (\ref{effmass}), with this parametrisation 
is still consistent with the empirical 
knowledge of kaon-nucleus scattering and allows to explore 
in-medium effects on the production 
mechanism arising from zero temperature kaon potentials. 
For the kaon production via pion absorption $\pi B\longrightarrow YK^+ $ 
the elementary cross section of \cite{tuebingen1} are used. 
For the $N N \longrightarrow BYK^+ $ channels we apply 
the cross sections of Ref. \cite{sibirtsev95} which give a good fit 
to the COSY-data close to threshold. For the case of 
$N \Delta \longrightarrow BYK^+ $ and 
$\Delta \Delta \longrightarrow BYK^+ $ reactions experimental data 
are rare. Thus we rely on the model calculation of ref. \cite{tsushima99}. 
In the case that 
a $N^*$ resonance is involved in the reaction we used the same 
cross section as for nucleons. 
In the presence of scalar and vector fields the kaon optical potential 
in nuclear matter has the same structure as the corresponding 
Schroedinger equivalent optical potential for nucleons 
\beq
U_{\rm opt}(\rho ,{\bf k}) 
=  -\Sigma_S + \frac{1}{m_{\mathrm K}} k_{\mu} V^{\mu}  
+ \frac{\Sigma_S^2 - V_{\mu}^2}{2m_{\mathrm K}}  ~.
\label{uopt}
\end{equation}
and leads to a shift of the thresholds conditions inside 
the medium. To fulfill 
energy-momentum conservation the optical potential is absorbed 
into an newly defined effective mass 
\beq
{\tilde m}_{\mathrm K} (\rho ,{\bf k}) 
= \sqrt{ m_{\mathrm K}^2 + 2m_{\mathrm K} U_{\rm opt}(\rho ,{\bf k}) }
\label{effmass}
\eeq
which is a Lorentz scalar and sets the canonical momenta on the mass-shell 
$0=  k_{\mu}^{2} - {\tilde m}_{\mathrm K}^{2}$. 
Thus, e.g., the threshold condition for $K^+$ production in baryon induced 
reactions reads 
$\sqrt{s} \ge {\tilde m}_B + {\tilde m}_Y + {\tilde m}_K$ 
with $\sqrt{s}$ the center--of--mass energy of the colliding baryons. 
For a consistent treatment of the thresholds the scalar 
and vector baryon mean fields entering into eq. 
(\ref{effmass}) are determined from two 
versions of the non-linear 
Walecka model with K=200/380 MeV, respectively \cite{li95b}. The hyperon 
field is thereby scaled by 2/3 which yields also a 
good description of the $\Lambda$ flow \cite{wang99b}. Since the 
parameterizations chosen for the non-linear Walecka model yield 
the same EOS as the Skyrme ones, the overall energy is conserved. 
The kaon production is treated perturbatively and does generally 
not affect the reaction dynamics \cite{fang94}.
\section{Probing the nuclear EOS by subthreshold $K^+$ production}
The $K^+$ excitation function for $Au+Au$ and $C+C$ 
reactions starting from 0.8 A$\cdot$GeV which is far below 
threshold ($E_{thr}=$ 1.58 GeV) has been measured by the KaoS 
Collaboration \cite{sturm00,kaos99}. In \cite{fuchs01} we 
calculated this excitation function for a soft/hard EOS 
including the in-medium kaon potential. 
For both systems the agreement with the 
KaoS data \cite{sturm00} is very good when a soft EOS is used. 
In the large system there was 
a visible EOS effect which is absent in the light system. 
The inclusion of the repulsive in-medium $K^+$ potential is 
thereby essential to reproduce the data \cite{kaos99}.  
Already in the light system the $K^+$ yield is reduced by about 
$50\%$. To extract more clear information on the nuclear EOS,  
in Fig. 1 we considered the ratio $R$ of the 
kaon multiplicities obtained in $Au+Au$ over $C+C$ 
reactions, normalized to the corresponding mass numbers.
The kaon potential is included since without the in-medium potential 
one is not able to reproduce the experimental $K^+$ yields 
\cite{fuchs01,hartnack01}. 
\begin{figure}[h]
\begin{minipage}[h]{95mm}
\unitlength1cm
\begin{picture}(9.,9.5)
\put(0.5,0){\makebox{\epsfig{file=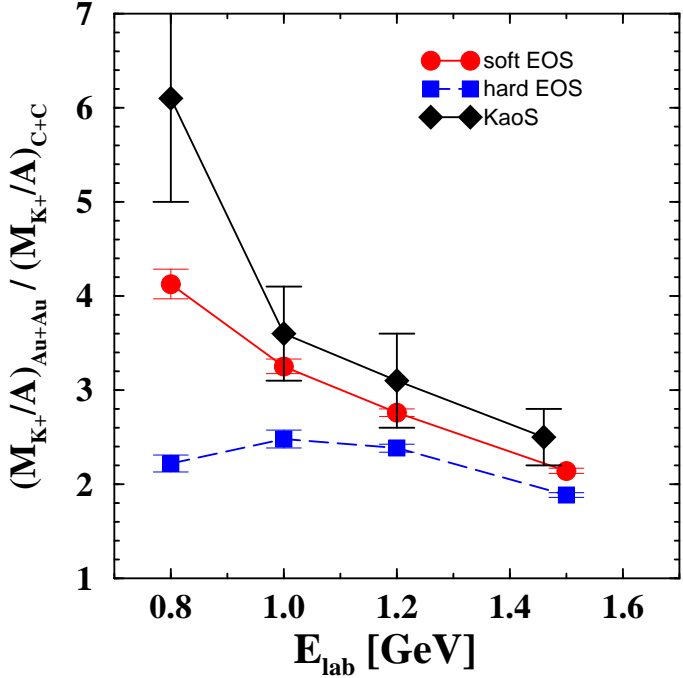,width=9.0cm}}}
\end{picture}
\end{minipage}
\vspace*{-1.5cm}
\hspace*{1.0cm}
\begin{minipage}[h]{70mm}
\caption{Excitation function of the ratio $R$ of $K^+$ 
multiplicities obtained in inclusive $Au+Au$ over $C+C$ 
reactions. The calculations are performed with 
in-medium kaon potential and using a hard/soft nuclear EOS and 
compared to the data from the KaoS Collaboration \protect\cite{sturm00}. 
}
\end{minipage}
\label{Fig1}
\vspace*{1.5cm}
\end{figure}
The calculations are performed under 
minimal bias conditions with $b_{{\rm max}}=11$ fm 
for $Au+Au$ and $b_{{\rm max}}=5$ fm for $C+C$ and normalized to the 
experimental reaction cross sections \cite{sturm00,kaos99}. Both 
calculations show an increase of $R$ with decreasing incident energy 
down to 1.0 A$\cdot$GeV. However, this increase is much less 
pronounced when the stiff EOS is employed. 
In the latter  case $R$ even decreases at 0.8 A$\cdot$GeV 
whereas the soft EOS leads to an unrelieved increase of $R$. 
At 1.5 A$\cdot$GeV which is already very close to threshold 
the differences between the two models become small. 
The strong increase of $R$ can be directly related to 
higher compressible nuclear matter. The comparison to the experimental 
data from KaoS \cite{sturm00} where the increase of $R$ is even more 
pronounced strongly favors a soft equation of state. 

To obtain a quantitative picture of the explored density 
effects in Fig. 2 the baryon densities are shown at 
which the kaons are created. The energy is chosen most below 
threshold, i.e. at 0.8 A$\cdot$GeV and only central collisions 
are considered where the effects are maximal. $dM_{K^+}/d\rho$ 
is defined as 
\beq
dM_{K^+}/d\rho = \sum_{i}^{N_{K^+}} 
\frac{d P_i }{ d\rho_B ({\bf x}_i, t_i)}
\eeq 
where $\rho_B$ is the baryon density at which the kaon $i$ was created 
and  $P_i$ is the corresponding production probability. For 
the comparison of the two systems the curves are normalized to the 
corresponding mass numbers. 
\begin{figure}[h]
\begin{minipage}[h]{95mm}
\unitlength1cm
\begin{picture}(9.,9.0)
\put(0.5,0){\makebox{\epsfig{file=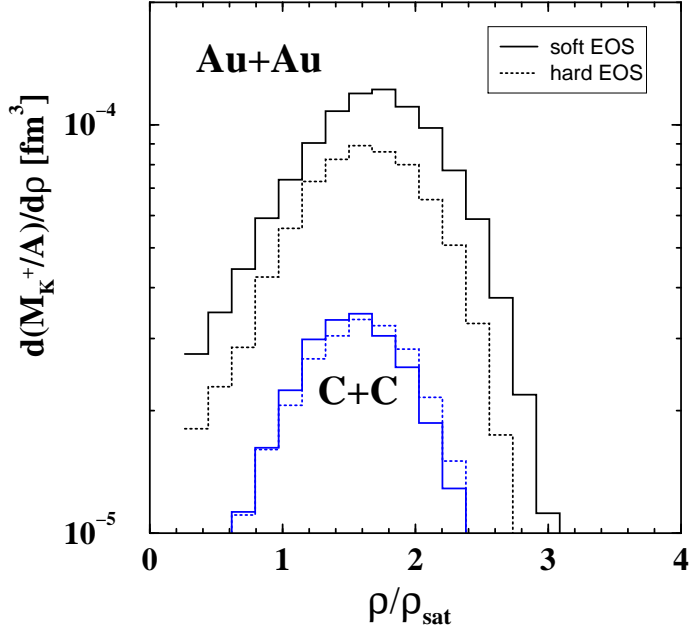,width=9.0cm}}}
\end{picture}
\end{minipage}
\vspace*{-1.5cm}
\hspace*{1.0cm}
\begin{minipage}[h]{70mm}
\caption{Kaon multiplicities (normalized to the mass numbers of the 
colliding nuclei) as a function of the baryon density 
at the space-time coordinates 
where the $K^+$ mesons have been created. Central (b=0 fm) $Au+Au$ and $C+C$ 
reactions at 0.8 A$\cdot$GeV are considered. 
The calculations are performed with 
in-medium kaon potential and using a hard/soft nuclear EOS. 
}
\end{minipage}
\label{Fig3}
\vspace*{1.5cm}
\end{figure}
Fig.2 illustrates several features: 
Only in the case of a soft EOS the mean densities at which kaons 
are created differ significantly for the two different reaction 
systems, i.e. $<\rho /\rho_{\rm sat} >$=1.46/1.40 for $C+C$  
and 1.47/1.57 for $Au+Au$ using  
the hard/soft EOS. Generally, in $C+C$ reactions densities above 
$2\rho_{\rm sat}$ are rarely reached whereas in $Au+Au$ the kaons are 
created at densities up to three times saturation density. 
Furthermore, for $C+C$ the density distributions are weakly 
dependent on the nuclear EOS. The situation changes 
completely in $Au+Au$. Here the densities profile 
shows a pronounced EOS dependence \cite{li95b}. 
Moreover, the excess of kaons obtained with the soft EOS 
originates almost exclusively from 
high density matter which demonstrates that compression effects 
are probed. 
\begin{figure}[h]
\begin{minipage}[h]{95mm}
\unitlength1cm
\begin{picture}(9.5,9.0)
\put(0.5,0){\makebox{\epsfig{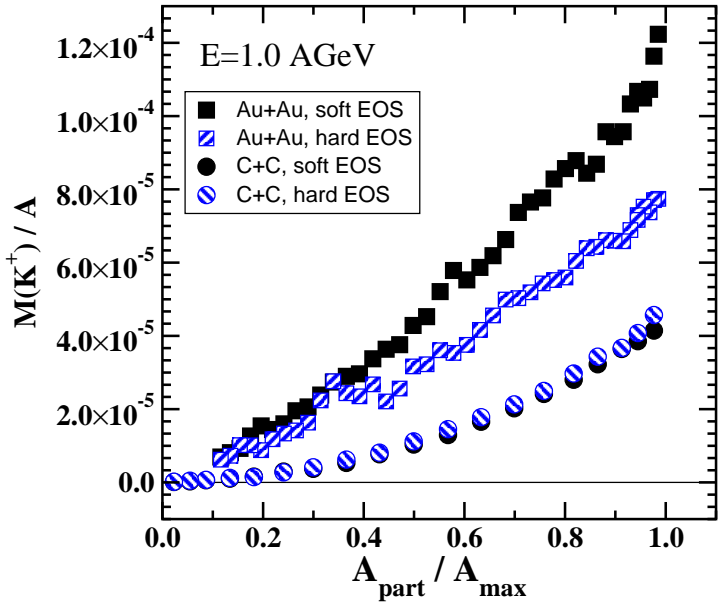}}}
\end{picture}
\end{minipage}
\vspace*{-1.5cm}
\hspace*{1.0cm}
\begin{minipage}[h]{70mm}
\caption{Kaon multiplicities as a function of $A_{\rm part}$ at 
E=1.0 AGeV. Multiplicities as well as $A_{\rm part}$ are 
normalized to the mass numbers of the 
colliding nuclei. 
}
\end{minipage}
\label{Fig_Apart}
\vspace*{1.5cm}
\end{figure}

The density effect is also clearly reflected in the $A_{\rm part}$ 
dependence of the kaon multiplicities. Fig. \ref {Fig_Apart} compares 
the $A_{\rm part}$ dependence in $Au+Au$ and $C+C$ reactions, now for 1.0 AGeV
laboratory energy. In both cases the kaon multiplicities as well as 
$A_{\rm part}$ which has been derived within the geometrical model, 
 are normalized to the corresponding mass numbers. 
As can be seen from there, in $C+C$ the $A_{\rm part}$ dependence of
the kaon production is completely insensitive to the nuclear equation of
state. The large system, in contrast, shows a distinct EOS dependence. 
In $Au+Au$ the enhanced kaon production which is due to higher 
compression when a soft EOS is used, 
becomes more and more pronounced with increasing
centrality. However, one has to keep in mind that the large difference 
between the soft and hard EOSs in most central reactions is washed out 
to some extent in minimal bias reactions. There the bulk of kaons 
originates from semi-central reactions $b\sim 5$ fm, corresponding 
to $A_{\rm part}/A_{\rm max}\sim 0.7$. In this context it will be 
very helpful to study the EOS dependence also in $C+Au$ reactions 
and to compare to forthcoming data from KaoS \cite{uhlig03}.

\subsection{How firm are the conclusions?}
Of course now the question arises, how firm conclusions on the
nuclear EOS are which can be drawn from kaon production in heavy 
ion reactions. Possible concerns might be 
based on the facts that subthreshold kaons are an extremely rare 
probe and not all elementary production cross sections have been 
measured. 

In heavy ion collisions the $K^+$ production runs over two major 
channels, namely baryon-baryon induced reactions 
$BB\mapsto BYK^+$ and pion-baryon induced reactions $\pi B\mapsto
YK^+$ which are both about equally important \cite{fuchs97}. In 
both cases the initial baryons can either be nucleons or nucleon 
resonances (mainly $\Delta (1232)$), the hyperons are $\Lambda$ or 
$\Sigma$ hyperons. Processes  with nucleon resonances in the final 
state are energetically suppressed. Concerning the knowledge 
of the elementary reaction cross section the situation is presently 
as follows: the $NN$ and $ \pi N$ cross sections are quite well under
control since these channels have been measured in $pp$ reactions 
\cite{cosy11} and in $\pi^\pm p$ reactions. The reactions which 
involve nucleon resonances, in particular with $\Delta$'s in the 
initial states ($ i= N\Delta, \pi\Delta, \Delta\Delta$) are less secure 
due to the lack of corresponding experimental data. 
Thus one has to rely on model assumptions. 
The cross sections which have been used in the present transport 
calculations are based on the effective Lagrangian model of 
Refs. \cite{tsushima99,tuebingen1,tsushima00}. The isospin dependence 
of the cross sections is thereby determined in the standard way 
by isotopic relations assuming iso-spin independent matrix elements.   

\begin{figure}[h]
\unitlength1cm
\begin{picture}(12.,9)
\put(1.5,0){\makebox{\epsfig{file=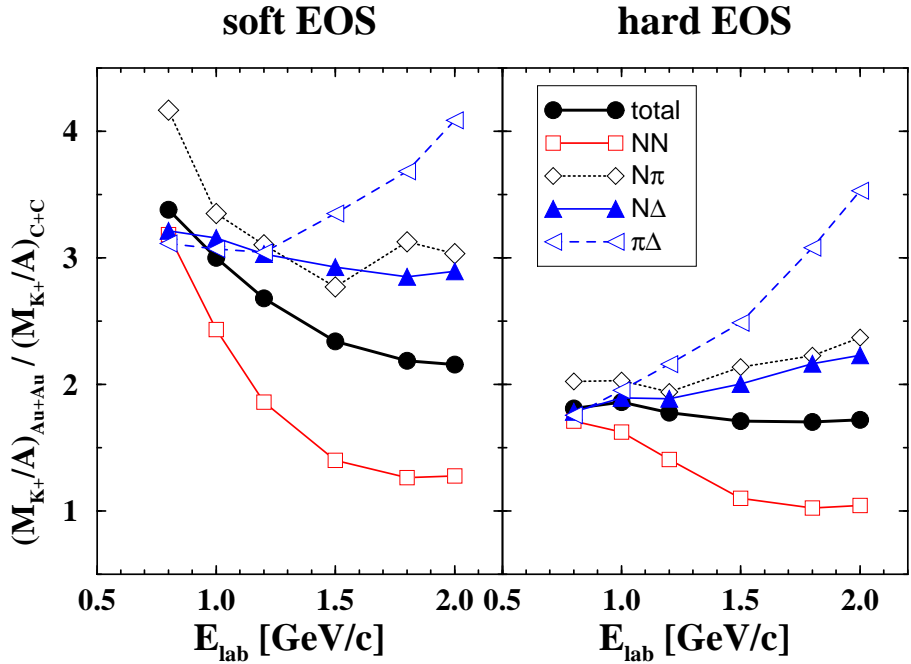,width=12.0cm}}}
\end{picture}
\caption{Dependence of the excitation function of $R$ on the 
various $K^+$ production channels. Central (b=0 fm) $Au+Au$ and $C+C$ 
reactions are considered. The calculations are performed with 
in-medium kaon potential.
}
\label{fig_chan}
\end{figure}

Hence there exists still some uncertainty in the transport
calculations due to the incomplete knowledge of the elementary 
reaction cross sections. There are, however, two good arguments 
why conclusions on the EOS dependence of the kaon production 
should be rather robust against such possible uncertainties:
\begin{itemize}
\item Changes of the production cross sections shift absolute yields   
but considering the ratio of different reaction systems such errors 
drop out in leading order.
\item Conclusions are based on the slope of this ratio as a function
of energy. It is rather unlikely that an incomplete knowledge 
of the cross sections, e.g. concerning their isospin dependence, 
can create the observed energy dependence. The systematics of spurious
contributions should be flat as a function of energy. 
Otherwise one have to assume extremely unconventional threshold 
effects. 
\end{itemize}
To be more quantitative we consider in the next figure the 
excitation function of $R$ for the various production channels. 
There the ratios $R_i$ are built separately for the production 
channels with initial states $ i= NN, \pi N, N\Delta, \pi\Delta, 
\Delta\Delta$.
As can be seen from Fig. \ref{fig_chan}, 
the shape of $R$ is not strongly influenced 
by the $N\Delta~, \pi\Delta$ channels which are the most insecure ones. 
The excitation function for 
the $N\Delta$ contribution varies only little as a function of energy 
and is similar using the different EOSs. The contribution of the 
$\pi\Delta$ channel is decreasing for 
both, a hard and a soft EOS. The shape of $R$ is to most extent 
determined by the $NN$ and $ \pi N$ contributions. In our calculations 
the latter channel is responsible for the decrease of $R$ very far 
below threshold when the hard EOS is applied. 
These findings are generally confirmed by 
independent transport calculations of 
the Nantes group using the IQMD transport model \cite{hartnack01} 
shwon in Fig. 5.
These calculations include an in-medium kaon potential derived in 
relativistic mean field theory (RMF) \cite{schaffner97} 
which is somewhat less repulsive than that one 
used in our calculations. For the soft 
EOS the IQMD calculations coincide almost with the present 
results \cite{fuchs01}. For the hard EOS there exist still deviations 
concerning the slope of $R$ going far below threshold. This could be due 
to the different in-medium potentials and is an open question which 
has to be resolved by future investigations. However, the two 
sets of transport calculations show a good overall agreement and 
both rule out the hard EOS from the comparison with data. The 
shaded area in Fig. 5 can be taken as the existing range of 
uncertainty in the theoretical model description of the considered 
observable. 

\begin{figure}[h]
\begin{minipage}[h]{95mm}
\unitlength1cm
\begin{picture}(9.5,9.0)
\put(0.5,0){\makebox{\epsfig{file=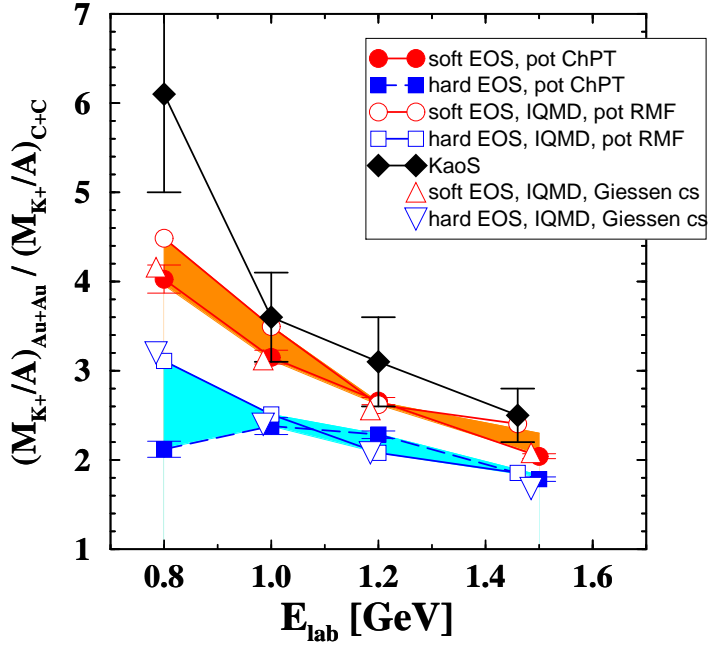,width=9.5cm}}}
\end{picture}
\end{minipage}
\vspace*{-1.5cm}
\hspace*{1.0cm}
\begin{minipage}[h]{70mm}
\caption{Excitation function of the ratio $R$ of $K^+$ 
multiplicities obtained in inclusive $Au+Au$ over $C+C$ 
reactions. Our results are compared to independent IQMD 
calculations \protect\cite{hartnack01}. The shaded area indicates
thereby the range of uncertainty in the theoretical models. 
In addition IQMD results based on an alternative set of 
elementary $K^+$ production cross sections are shown.
}
\end{minipage}
\label{FigIQMD}
\vspace*{1.5cm}
\end{figure}
Moreover, the IQMD calculations were also repeated with an alternative 
set of $N\Delta; \Delta\Delta \mapsto NYK^+$ cross sections taken from
\cite{cassing99} which are about a factor of two larger than those  
from Tsushima et al. \cite{tsushima99}. The ratio $R$ is almost 
completely independent on this change. Another prove for the
robustness of this observable is the fact influence of the repulsive 
$K^+$ potential which decreases the total kaon yield by about a factor 
drops almost completely out when ratio of the two different mass 
systems is built \cite{fuchs01,hartnack01}. 

\subsection{Are the conclusions consistent with 
information from other sources?}
Finally the question arises up to which degree a consistent picture 
has emerged after more than ten years of intensive experimental and 
theoretical efforts to understand the kaon production at 
intermediate energies. In the following I will argue that 
concerning $K^+$ we have in the meantime a rather consistent 
picture while for $K^-$ the situation is not yet so clear. 

The reason is that subthreshold $K^+$ production is easier to handle, 
both from the  experimental and the theoretical side. Due to the 
lower threshold there are much more data with much higher 
precision available than for $K^-$. Also theoretically the medium 
dependence of the $K^+$ meson properties are better under
control. The mean field approximation seems to work, i.e. mass shifts 
and can be taken from the leading order chiral Lagrangian
\cite{kapla86,brown962}  and in the medium there exits still a 
well established quasi-particle pole \cite{waas97,lutz98}. This allows a 
treatment within standard transport based on the quasi-particle 
approximation. The $K^- -N$ system, on the other hand, lies in the 
vicinity of the $\Lambda_{1405}$ resonance which implies a strong
coupling to this state. Hence a simple mean field picture will not 
work but a coupled channel treatment of the $K^-$ in the medium 
is necessary \cite{waas97,lutz02,tolos03}. As a consequence, the $K^-$ has 
complicated spectral properties in the medium and a simple 
quasi-particle picture is nor more suitable \cite{lutz02,tolos03}. This 
requires a more sophisticated treatment within transport simulations 
which accounts at least approximately for the off-shell contributions 
of the spectral functions \cite{cassing03}. In the latter case 
the interpretation of existing data with the help of transport
simulation has not yet reached a level which allows to say that the 
 $K^-$ properties are settled.    

Coming back to the $K^+$ the situation is much more satisfying. 
Most transport calculations agree on the necessity of a repulsive  $K^+$ 
potential in order to understand total yields as well as the collective 
motion of  $K^+$ mesons, i.e. in-plane and out of-plane flow 
\cite{fopi95,li97,wang97,fuchs98,brat97,nantes99}. This picture was recently
complemented by measurements of the $K^{+}$ production in proton-nucleus 
reactions \cite{anke}. Although such reactions test only subnormal 
nuclear densities they are much easier to handle than the complicated 
dynamical evolution of heavy ion reactions. Corresponding data $pA$ 
from ANKE revealed strong evidence for a repulsive  $K^+$ 
potential which is of the order of magnitude as predicted by 
effective chiral lagrangiens. 
\subsubsection{Many-body calculations}
Concerning the nuclear equation of state one has to confront the 
information from subthreshold $K^+$ production with the knowledge 
obtained from other sources: 
At intermediate energies heavy ion reactions test the density 
range between two 
and three times nuclear density. The information from kaon 
production implies  
that in this density range the EOS should show a soft behavior. 
One has of course to be aware that the adopted Skyrme forces are 
simplified interactions which are easy to handle but must not be 
very realistic. 
\begin{figure}[h]
\begin{minipage}[h]{100mm}
\unitlength1cm
\begin{picture}(10.,8.0)
\put(0.5,0){\makebox{\epsfig{file=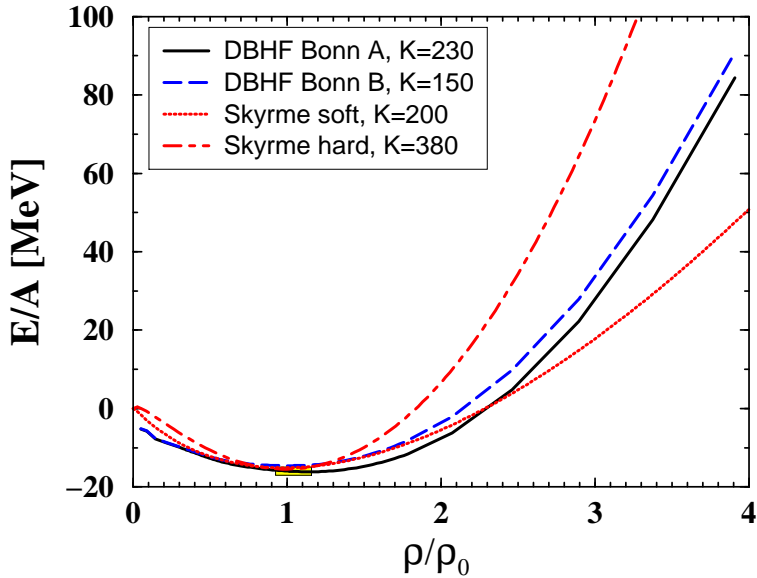,width=10cm}}}
\end{picture}
\end{minipage}
\vspace*{-1.5cm}
\hspace*{1.0cm}
\begin{minipage}[h]{60mm}
\caption{The Nuclear matter EOS from microscopic
Dirac-Brueckner-Hartree-Fock calculations from
\protect\cite{boelting99} are compared to that  
resulting from soft/hard Skyrme forces. 
}
\end{minipage}
\label{FigDB}
\vspace*{1.5cm}
\end{figure}
A microscopic approach to nuclear matter which is based on realistic 
nucleon-nucleon interactions has to solve the correlated quantum-mechanical
many-body problem. Such an approach is ,e.g., the Brueckner-Hartree-Fock 
approach which accounts for the two-body correlation in the medium solving 
the Bethe-Goldstone equation, i.e. the Lippmann-Schwinger equation in
the medium. The relativistic version, i.e. the
Dirac-Brueckner-Hartree-Fock (DBHF) 
approach delivers quite reasonable values for the nuclear saturation 
mechanism \cite{dbhf} and can thus be considered as a reliable 
method to extrapolate the many-body approach to higher nuclear 
densities. In Fig. 6 we compare 
DBHF results obtained with two slightly different NN-interactions, 
i.e. Bonn A and Bonn B \cite{bonn} to the simple Skyrme type 
equations of state. The results of advanced DBHF calculations 
are taken from \cite{boelting99}. In this context it may worthwhile 
to mention that very similar results to \cite{boelting99} were 
recently obtained by Weise and coworkers in a treatment of 
the  nuclear many-body problem based on chiral perturbation theory 
\cite{finelli}. Generally such many-body calculations predict 
a relatively soft behavior of the EOS in the relevant density range, 
i.e. for densities below about three times saturation density. 
In the microscopic 
approach the high density behavior is, however, only loosely connected 
to the curvature at saturation density. For Bonn A and B the 
compressibilities are quite different while the EOSs at high 
densities are very close. Below $3\rho_0$ both are not too far 
from the soft Skyrme EOS.
\subsubsection{Nucleon Flow}
Another observable which helps to constrain the nuclear mean field 
and the underlying EOS at supra-normal densities 
is the collective nucleon flow \cite{norbert99}. 
The transverse flow $v_1$ has been found to be sensitive to the EOS
and, in particular in peripheral reactions, to the momentum dependence 
of the mean field \cite{dani00,gaitanos01}. The elliptic flow $v_2$, 
in addition, is very sensitive to the maximal compression reached in
the early phase of a heavy ion reaction. The cross over from
preferential in-plane flow $v_2 < 0$ to preferential 
out-off-plane flow  $v_2 > 0$ around 4-6 AGeV has also led to
speculations about a phase transition in this energy region which 
goes along with a softening of the EOS \cite{eflow}.  

The present situation can be summarized as follows: At SIS energies 
existing flow data are consistent with the usage of mean fields 
which are close to those obtained from microscopic DBHF calculations, 
both concerning their density and momentum dependence
\cite{dani00,gaitanos01}. A detailed comparison to FOPI data 
\cite{andronic99} for $v_1$ and $v_2$ between 0.2 and 0.8 AGeV favors 
thereby a relatively soft EOS \cite{gaitanos01} such as the DBHF 
result for Bonn A (K=230 MeV, shown in Figure 6). 
The full flow excitation function, ranging from low SIS up to top 
AGS energies, has been studied in \cite{dani02}. The conclusion from 
this study was that, both, super-soft equations of state (K=167 MeV) 
as well as hard EOSs (K$>$300 MeV) are ruled out by  data. Hence the 
picture is again consistent with the information obtained from 
kaon production.

\section{Summary}
To summarize, we find that at incident energies far below 
the free threshold $K^+$ production is a suitable tool to study the 
dependence on the nuclear equation of state. Using a light system 
as reference frame there is a visible sensitivity on the EOS 
when ratios of heavy ($Au+Au$) over light ($C+C$) systems are considered. 
Transport calculations indicate that the $K^+$ production gets 
hardly affected by compressional effects in $C+C$ but is 
highly sensitive to the high density matter 
($1\le \rho /\rho_{\rm sat} \le 3$) created in $Au+Au$ reactions. 
Results for the  $K^+$ excitation function in $Au+Au$ over 
$C+C$ reactions as measured by the KaoS Collaboration, strongly 
support the scenario of a soft EOS. This statement turns out to 
be rather robust against possible model uncertainties:  
It is almost independent on the variation of particular reaction
channels where elementary cross sections are uncertain. It is also 
insensitive to 
the inclusion/neglection of a changing in-medium kaon mass as predicted 
by chiral models. The idea of a soft EOS in the considered density 
range is also consistent with the knowledge from microscopic many-body 
theory and from nuclear flow analysis in heavy ion reactions. 

Concerning the quest for in-medium modification of the kaon 
properties transport calculation have in the meantime reached a 
certain level of consistency: The explanation 
of the total $K^+$ yields and the  $K^+$ flow requires 
the presence of a repulsive in-medium potential.  This picture has been  
complemented by measurements of kaon production in $p+A$ reactions.
\subsection*{Acknowledgments}
I would like to thank the following persons who all contributed to the 
results discussed in the present work: Amand Faessler,
T. Gross-Boelting, Z. Wang, E. Zabrodin, and Y.-M. Zheng.


\end{document}